\documentclass[prd,amsmath,amssymb]{revtex4}

\usepackage{graphicx}
\usepackage{dcolumn}
\usepackage{bm}
\usepackage{cases}
\usepackage{mathtools}

\begin{document}

\title{Local symmetries, anomalies and constrains in Burgers Turbulence\\ }

\author{Timo\hspace*{1mm}Aukusti\hspace*{1mm}Laine \vspace*{0.5cm}}
\email{timo.au.laine@gmail.com}



           
\begin{abstract}
We study stochastic Burgers turbulence without pressure. We first show that the variational derivative of the Burgers equation is dependent on the velocity field, suggesting the existence of an anomaly. The anomaly is created by an operator that is non-self-adjoint. To calculate it correctly, we need to find its square. There are similarities with conformal and chiral two-dimensional field theories, but causality is the key that makes the difference. We calculate the determinant and use two local symmetries to verify the result. By requiring the disappearance of the anomaly, the velocity field is constrained and the theory becomes anomaly-free. These symmetries obey Kolmogorov's second law of self-similarity. One can choose an anomaly-free theory, a partially broken theory, or a fully broken theory by choosing the constraint appropriately. There is an analogy to gauge fixing or vacuum selection which define the local configuration.

\end{abstract}

\maketitle

\vspace*{0.5cm}

\section{\label{sec0}Introduction}

Turbulence, characterized by its seemingly random and complex patterns of motion, remains one of the most fascinating and challenging phenomena in physics. Its impact spans across diverse disciplines, from engineering and environmental sciences to astrophysics and meteorology. Despite its ubiquity, turbulence continues to defy complete understanding, making it a subject of perpetual fascination and exploration.

Burgers turbulence provides an elegant and insightful approach to study turbulence without the full complexity of the Navier-Stokes equations governing fluid motion. In this model, the nonlinear advection term is retained while the pressure gradient term is neglected, making it more tractable for analysis. As a result, Burgers turbulence presents an ideal testing ground for developing theories, numerical simulations, and experimental investigations to comprehend the fundamental characteristics of turbulent flows.

In this article we apply field theory to Burgers turbulence. We show the local symmetry which is constrained by the anomaly. Some of the results were already shown in Ref.~\onlinecite{laine}, but here we provide more details and new results. The generalization to the stochastic Navier-Stokes turbulence is shown in Ref.~\onlinecite{laine2}.

\section{\label{sec1}Burgers turbulence}

We investigate one-dimensional Burgers equation without pressure 

\begin{equation}
  u_t + uu_x -\nu u_{xx} = f(t,x), \label{eq1}
\end{equation}

\noindent
where $u = u(t,x)$ is a velocity field, $\nu$ is a viscosity, and $f(t,x)$  is a Gaussian random force,

\begin{equation}
  \langle f(t,x)f(t',y) \rangle = \kappa(x-y)\delta(t-t'). \label{eq2}
\end{equation}

\noindent
Function $\kappa(x-y)$ defines the spatial correlation of the random forces. This equation has been studied extensively, for example, in Refs.~\onlinecite{polyakov}-\onlinecite{ivaskevich}.

We write equations (\ref{eq1})-(\ref{eq2}) as

\begin{equation}
  \langle F[\lambda]\rangle = \int D\mu Du F[\lambda] J[u]\exp(-S[u,\mu]),  \label{eq:S0}
\end{equation}

\noindent
where the action $S[u(t,x),\mu(t,x)]$ is defined as

\begin{equation}
  S[u,\mu] = \frac{1}{2} \int dtdxdy \mu(t,x)\kappa(x-y)\mu(t,y) -i\int
dtdx \mu(u_t+uu_x-\nu u_{xx}),  \label{eq:S}
\end{equation}

\noindent
and the Jacobian is

\begin{equation}
  J[u] = \det \biggl | \frac{\delta f}{\delta u}\biggr | = \det | \partial_t+\frac{\delta f_1}{\delta u} | =
  \det | \partial_t+u_x+u\partial_x-\nu \partial_{xx} |.
\label{eq:jac}
\end{equation}

\noindent
 By using anticommuting functions, $\Psi = \Psi(t,x,u)$ and $\tilde{\Psi} = \tilde{\Psi}(t,x,u)$, Refs.~\onlinecite{zinn}-\onlinecite{ramond},  the determinant can be written as

\begin{equation}
  J[u] = \int D\bar{\Psi}D\Psi \exp(-S_A),
\end{equation}

\noindent
where the action of the determinant is

\begin{equation}
  S_A = -\int dtdx \bar{\Psi}(\partial_t+u_x+u\partial_x-\nu \partial_{xx})\Psi. \label{det1}
\end{equation}

\noindent
The complete action in (\ref{eq:S0}) takes the form

\begin{eqnarray}
  S_{tot}[u,\mu,\Psi,\tilde{\Psi}] &=& \frac{1}{2} \int dtdxdy \mu(t,x)\kappa(x-y)\mu(t,y) -i\int
dtdx \mu(u_t+uu_x-\nu u_{xx}) \nonumber \\
 && -\int dtdx \bar{\Psi}(\partial_t+u_x+u\partial_x-\nu \partial_{xx})\Psi.
 \label{eq:Sf}
\end{eqnarray}

\noindent
Action (\ref{eq:Sf}) has some interesting characteristics that we will examine.

In this article, we focus on studying the properties of the determinant (\ref{eq:jac}). In particular, we show that it is not a number and there is physics involved. This can be shown by, $\nu = 0$,

\begin{equation}
  \frac{\delta S_A}{\delta u} =  \bar{\Psi}_x\Psi - \frac{\delta ( \bar{\Psi}\Psi_t) }{\delta u}+ u\frac{\delta ( \bar{\Psi}_x\Psi )}{\delta u}. \label{det222}
\end{equation}

\noindent
If we assume that the fields $\Psi$ and $\bar{\Psi}$ are independent of $u$, the variation (\ref{det222}) still does not vanish. Alternatively, if the variation (\ref{det222}) is claimed to be zero, the field $u$ has a constraint. This shows that the action (\ref{det1}), and therefore also the determinant (\ref{eq:jac}), depends on the field $u$. In this article we calculate the determinant.
There are striking similarities with, for example, conformal field theory, but in turbulence the requirement of causality makes the equations different.\\

\section{\label{sec2}Determinant}

In this section we study the local symmetries of the determinant action (\ref{det1}) and calculate the Jacobian.  First we observe that

\begin{equation}
  \int dtdx \bar{\Psi}(\partial_t+u_x+u\partial_x-\nu \partial_{xx})\Psi
  = \int dtdx (-\partial_t-u\partial_x-\nu \partial_{xx})\bar{\Psi}\Psi. \label{det123}
\end{equation}

\noindent
Therefore the determinant (\ref{eq:jac}) has two representations

\begin{equation}
  S_A = -\int dtdx \bar{\Psi}(\partial_t+u_x+u\partial_x-\nu \partial_{xx})\Psi, \label{det1a}
\end{equation}

\noindent
and

\begin{equation}
  S_B = -\int dtdx \bar{\Psi}(-\partial_t-u\partial_x-\nu \partial_{xx})\Psi. \label{det1b}
\end{equation}

\noindent
We call (\ref{det1a}) the "right-handed" determinant and (\ref{det1b}) the "left-handed" determinant. Next we study the local symmetries of these determinants. We first set the viscosity to zero. \\

\noindent
{\bf Right-handed determinant} \\

There are two local symmetries. The action 

\begin{equation}
  S_A(\nu = 0) = -\int dtdx \bar{\Psi}(\partial_t+u_x+u\partial_x)\Psi \label{det1a1}
\end{equation}

\noindent
is  invariant to  local time reparameterization, $\alpha = \alpha(t)$,

\begin{eqnarray}
  &&\delta u_{A1} = \alpha u_t+ \alpha'u, \label{detu} \\
  &&\delta \Psi_{A1} = \alpha \Psi_t,\\
  &&\delta \bar{\Psi}_{A1} = \alpha \bar{\Psi}_t.
\end{eqnarray} 

\noindent
The action (\ref{det1a1}) is also invariant to local space-time symmetry, $\epsilon = \epsilon(t,x)$, 

\begin{eqnarray}
  &&\delta u_{A2} = \epsilon u_x-\epsilon_x u -\epsilon_t, \label{detpsi1}\\
  &&\delta \Psi_{A2} = \epsilon \Psi_x+\epsilon_x\Psi, \\
  &&\delta \bar{\Psi}_{A2} = \epsilon \bar{\Psi}_x.\label{detpsi2}
\end{eqnarray}

\noindent
{\bf Left-handed determinant} \\

In a same way, the action (\ref{det1b})

\begin{equation}
  S_B(\nu = 0) = -\int dtdx \bar{\Psi}(-\partial_t-u\partial_x)\Psi \label{det1b1}
\end{equation}

\noindent
is  invariant to  local time reparameterization, $\alpha = \alpha(t)$,

\begin{eqnarray}
  &&\delta u_{B1} = \alpha u_t+ \alpha'u, \label{detub} \\
  &&\delta \Psi_{B1} = \alpha \Psi_t,\\
  &&\delta \bar{\Psi}_{B1} = \alpha \bar{\Psi}_t,
\end{eqnarray} 

\noindent
and a local space-time symmetry, $\epsilon = \epsilon(t,x)$,

\begin{eqnarray}
  &&\delta u_{B2} = \epsilon u_x-\epsilon_x u +\epsilon_t, \label{detpsib1} \\
  &&\delta \Psi_{B2} = \epsilon \Psi_x, \\
  &&\delta \bar{\Psi}_{B2} = \epsilon \bar{\Psi}_x
  +\epsilon_x\bar{\Psi} .\label{detpsib2}
\end{eqnarray} 

\noindent
As shown by (\ref{detpsi1})-(\ref{detpsi2}) and (\ref{detpsib1})-(\ref{detpsib2}), actions (\ref{det1a1}) and (\ref{det1b1}) obey different local transformations and therefore the corresponding determinants are different. 
The reason for the ambiguity is that in Jacobian (\ref{eq:jac}) operator $\delta f_1/\delta u$ is non-self-adjoint. There is an analogy to chiral field theories with right-handed and left-handed operators.
To calculate the determinant correctly, we need to square it. However, in the current situation the usual methods of two-dimensional field theories cannot be used. The key is the causality which must be preserved during the calculation. There are no imaginary values or dimensions, and all the values and functions must be real.\\

\noindent
{\bf Square of the determinant} \\

We start with definitions. By definition, a determinant is the product of its eigenvalues. In this case, the determinant is an operator $ \delta f/\delta u$  which is also a function. A way to calculate the determinant is to consider the following eigenvalue equations

\begin{equation}
  \begin{cases}
(\partial_t+u_x+u\partial_x )A=\lambda A, \\ 
(-\partial_t-u\partial_x )A=\lambda A,
  \end{cases} \label{eigeneq}
\end{equation}

\noindent
where $\lambda = \lambda(t,x,u)$ is an eigenvalue and $A = A(t,x,u)$ is the corresponding eigenfunction. For consistency, both determinants must produce the same eigenvalue functions. Putting the equations (\ref{eigeneq}) together gives

\begin{equation}
  \lambda = \frac{1}{2}u_x(t,x). \label{eigenval}
\end{equation} 

\noindent
By analogy with chiral theories, the result is already the "square root".
As the definition of the determinant states, the value of the determinant is then 

\begin{equation}
  \det | \partial_t+u_x+u\partial_x | = \frac{1}{2}u_x(t,x).
\label{eq:jac3}
\end{equation}

\noindent
When using the result, the determinant action can be written as

\begin{equation}
   S_{D}(\nu = 0) = -  \int dtdx \bar{\Psi} (t,x) \frac{1}{2}u_x(t,x)  \Psi (t,x) . \label{det2D}
\end{equation}

\noindent
We check the result by investigating the local symmetries of the action (\ref{det2D}). The action is  invariant to  local time reparameterization, $\alpha = \alpha(t)$,

\begin{eqnarray}
  &&\delta u_{D1} = \alpha u_t+ \alpha'u, \label{detd} \\
  &&\delta \Psi_{D1} = \alpha \Psi_t,\\
  &&\delta \bar{\Psi}_{D1} = \alpha \bar{\Psi}_t. \label{detd0} 
\end{eqnarray} 

\noindent
It is also invariant to a local space-time symmetry, $\epsilon = \epsilon(t,x)$,

\begin{eqnarray}
  &&\delta u_{D2} = \epsilon u_x-\epsilon_x u, \label{detd1}\\
  &&\delta \Psi_{D2} = \epsilon \Psi_x+\frac{\epsilon_x}{2}\Psi, \label{detd200}\\
  &&\delta \bar{\Psi}_{D2} = \epsilon \bar{\Psi}_x+\frac{\epsilon_x}{2}\bar{\Psi},\label{detd2}
\end{eqnarray} 

\noindent
when $\epsilon_{xx}=0$. This makes sense. Now the local space-time symmetry is symmetric with respect to the fields $\Psi$ and $\bar{\Psi}$, and it is a combination of right-handed and left-handed symmetries

\begin{eqnarray}
  &&\delta u_{D2} = \frac{1}{2}(\delta u_{A2}+\delta u_{B2}), \label{detd11}\\
  &&\delta \Psi_{D2} = \frac{1}{2}(\delta \Psi_{A2}+\delta \Psi_{B2}), \\
  &&\delta \bar{\Psi}_{D2} = \frac{1}{2}(\delta \bar{\Psi}_{A2}+\delta \bar{\Psi}_{B2}).\label{detd22}
\end{eqnarray} 

\noindent
We note that the symmetry transformation ({\ref{detd1}) is a 
Galilean transformation on $x$. \\

\noindent
{\bf Causal determinant} \\

The result (\ref{eq:jac3}) suggests that we should consider the following causal determinant

\begin{equation}
  \det | \partial_t+\frac{1}{2}u_x |.
\label{eq:jac4}
\end{equation}

\noindent
One may wonder how these determinants are similar or different. First we use symmetries to show the difference. Later we give another explanation.
When writing (\ref{eq:jac4}) with anticommuting fields, we get

\begin{equation}
   S_{Causal} = -  \int dtdx \bar{\Psi} (t,x)(\partial_t+ \frac{1}{2}u_x(t,x))  \Psi (t,x) . \label{det2D1}
\end{equation}

\noindent
The causal action (\ref{det2D1}) is invariant to  local time reparameterization, $\alpha = \alpha(t)$,

\begin{eqnarray}
  &&\delta u_{A1} = \alpha u_t+ \alpha'u, \label{detuaa} \\
  &&\delta \Psi_{A1} = \alpha \Psi_t,\\
  &&\delta \bar{\Psi}_{A1} = \alpha \bar{\Psi}_t.\label{detuaa1}
\end{eqnarray} 

\noindent
and it is also invariant to local space symmetry, $\epsilon = \epsilon(x)$, 

\begin{eqnarray}
  &&\delta u_{A2} = \epsilon u_x-\epsilon_x u \label{detpsi2aa0}\\
  &&\delta \Psi_{A2} = \epsilon \Psi_x+\epsilon_x\Psi, \label{detpsi2aa1} \\
  &&\delta \bar{\Psi}_{A2} = \epsilon \bar{\Psi}_x, \label{detpsi2aa}
\end{eqnarray} 

\noindent
 where $\epsilon_{xx}=0$ and $\epsilon_t = 0$. As one can verify, the transformations (\ref{detuaa}) and (\ref{detpsi2aa0}) are the same as (\ref{detd}) and (\ref{detd1}). However, the condition $\epsilon_t = 0$ makes a difference.
   This means that the symmetries (\ref{detuaa})-(\ref{detuaa1}) and (\ref{detpsi2aa0})-(\ref{detpsi2aa}) cannot be observed simultaneously. The original determinant (\ref{det1a1}) has two local symmetries that can be simultaneous. Although the equation is not a square, it is correct. If the determinant is replaced by a square with anticommuting fields, it must also have two concurrent local symmetries. We also note that the anticommuting fields (\ref{detpsi2aa1}) and (\ref{detpsi2aa}) are not symmetric with respect to each other, like  (\ref{detd200}) and (\ref{detd2}). We conclude that (\ref{det2D}) is the correct form of the determinant square of the stochastic Burgers equation.\\

\noindent
{\bf Constraint} \\

We have now the result of the determinant (\ref{eq:jac3}) 

\begin{equation}
  \det | \partial_t+u_x+u\partial_x | = \frac{1}{2}u_x(t,x),
\label{eq:jac3211}
\end{equation}

\noindent
and we should convert it to a path integral. There are examples of calculating the causal determinant of the Langevin equations, for example in Ref.~\onlinecite{zinn}, but here the situation is more complicated and the results are not directly applicable. We will show the reason later. The Langevin equations approach could be applied to the determinant (\ref{eq:jac4}).

First, it is noted that the requirement of the determinant to be constant gives a condition

\begin{equation}
  \det | \partial_t+u_x+u\partial_x | = \frac{1}{2}u_x(t,x) = {\rm constant}, 
\label{eq:jac3212}
\end{equation}

\noindent
or 

\begin{equation}
  \partial_x \delta u(t,x) = 0. \label{condiss}
\end{equation}

\noindent
So there is an anomaly (or constraint) in the determinant and in general the determinant is not a number. The constraint (\ref{condiss}) is the condition we expect from the anomaly. 
We use the identity

\begin{equation}
 {\rm det M} = \exp {{\rm Tr} [ {\rm ln M} } ],
\end{equation}

\noindent
where M is a matrix. The
 path-integral representation of the action is 

\begin{equation}
   S_{D}(\nu = 0) =    - \int dtdx \ln ( \frac{1}{2} u_x (t,x)), \label{det20100}
\end{equation}

\noindent
and the variation is then

\begin{equation}
   \delta S_D =   -2\int dtdx \frac{1}{u_x}\partial_x\delta u. \label{consti8}
\end{equation}

\noindent
This integral is zero if

\begin{eqnarray}
&& \delta u = 0, \hspace*{0.5cm}{\rm or}\\
&&\partial_x\delta u=0,\hspace*{0.5cm}{\rm or}\\
&& u_{xx}=0.
\end{eqnarray}

\noindent
The equation (\ref{consti8}) shows the constraint correctly.
\\

\noindent
{\bf Zero eigenvalue} \\

We have the eigenvalue equation (\ref{eigeneq})

\begin{equation}
  (\partial_t + u_x + u \partial_x) A(t,x) = \frac{1}{2}u_xA (t,x), \label{eigeq0}
\end{equation}

\noindent
which links the two determinants (\ref{det1a1}) and (\ref{det2D}) together. From a mathematical point of view, the determinants are different: the left side of (\ref{eigeq0}) contains a non-self-adjoint operator, while the right side of (\ref{eigeq0}) is self-adjoint. The right side (\ref{eigeq0}) is the "square root" of the left side. The symmetries of the determinants are different, (\ref{detpsi1})-(\ref{detpsi2}) and (\ref{detd1})-(\ref{detd2}). However, the eigenvalues and the eigenfunctions are the same as the equation (\ref{eigeq0})  states. This is the property that makes the determinants the same from a physical point of view.

We may now write (\ref{eigeq0})  as 

\begin{equation}
  (\partial_t + \frac{1}{2}u_x + u \partial_x) A(t,x) = 0\cdot A(t,x) = 0. \label{eigeq1}
\end{equation}

\noindent
This is a zero eigenvalue equation.
Zero eigenvalues are particularly interesting because they can enable additional physical properties, for example local symmetries, or enable the system to continuously change from one state to another. Here, the zero eigenvalue has many consequences; it makes two local symmetries simultaneous and visible in the stochastic Burgers equation.

The constraint (\ref{condiss}) also links to the zero eigenvalue equation (\ref{eigeq1}). It defines the situation when the eigenvalue equation is zero. Note that the constraint imposes no restrictions on $t$. Therefore we have

\begin{eqnarray}
  \partial_t A(t,x) &=&  0,\label{condit1}\\
  u\partial_x A(t,x) &=&  -\frac{1}{2}u_x(t,x)A(t,x).  \label{condit2} 
\end{eqnarray}

\noindent
The condition (\ref{condit1}) also shows why the Langevin equation approach to computing determinants, Ref.~\onlinecite{zinn}, does not work when calculating the determinant of the stochastic Burgers equation. The eigenfunction does not depend on time. We will use these values later when calculating the perturbation of the viscosity. We also note that the determinant (\ref{eq:jac4}) has different eigenvalues and therefore, for example, the condition (\ref{condit1}) does not apply to it.\\

\section{\label{sec3}Anomaly }

First, we show the effective action when the viscosity is set to zero, and then we calculate the anomaly when the viscosity term is added. \\

\noindent
{\bf Zero viscosity} \\

 We now collect the results and write the effective action as
 
 \begin{equation}
  S_{eff}[u,\mu,\nu = 0] =  \frac{1}{2} \int dtdxdy \mu(t,x)\kappa(x-y)\mu(t,y) -i\int
dtdx \mu(u_t+uu_x) -    \int dtdx \bar{\Psi}  \frac{1}{2}u_x \Psi, \label{acteff11000}
\end{equation}

\noindent
or

\begin{equation}
  S_{eff}[u,\mu,\nu = 0] =  \frac{1}{2} \int dtdxdy \mu(t,x)\kappa(x-y)\mu(t,y) -i\int
dtdx \mu(u_t+uu_x) -  \int dtdx \ln (   \frac{1}{2}u_x(t,x)) . \label{acteff110}
\end{equation}

\noindent
The determinant creates an anomaly in the action which is the last term in (\ref{acteff110}). The original determinant (\ref{det1a1}) has two local symmetries and also the square (or square root) of the determinant (\ref{det2D}) has two symmetries: (\ref{detd})-(\ref{detd0}) and ({\ref{detd1})-(\ref{detd2}). Therefore, we expect the anomaly to obey these two symmetries as well. We have also a constraint

\begin{equation}
 \partial_x \delta u(t,x) = 0,   \label{constr2}
\end{equation} 

\noindent
which is a condition for anomaly-free theory. \\

{\bf Non-zero viscosity} \\

Next we investigate the situation when the viscosity is non-zero. First we note that the viscosity term 

\begin{equation}
  S_{\nu}= i\int dtdx \bar{\Psi}(\partial_{xx})\Psi \label{det1a11}
\end{equation}

\noindent
is not invariant to the local symmetry transformation ({\ref{detd1})-(\ref{detd2})

\begin{equation}
  \delta S_{\nu} = i
  \int dtdx \bar{\Psi}(\frac{\epsilon_{xxx}}{2}+2\epsilon_{xx}\partial_x+2\epsilon_x\partial_{xx})\Psi.\label{det1a11b}
\end{equation}

\noindent
This means that, in general, the Burgers equation is not invariant in this symmetry.

We calculate the viscosity term. Now we consider the following eigenvalue equations

\begin{equation}
  \begin{cases}
(\partial_t+u_x+u\partial_x -  \nu \partial_{xx})A'=\lambda '  A', \\ 
(-\partial_t-u\partial_x -  \nu \partial_{xx})A'=\lambda ' A', \label{cases1}
  \end{cases}
\end{equation}

\noindent
where $\lambda ' = \lambda '(t,x,u)$ is the eigenvalue and $A '= A'(t,x,u)$ is the corresponding eigenfunction. This gives the result 

\begin{equation}
  \lambda ' = \frac{1}{2}u_x - \nu \frac{\partial_{xx}A'}{A'}
  = \lambda - \nu \frac{\partial_{xx}A'}{A'}
  \label{visc1}
\end{equation} 

\noindent
where $\lambda$ is the eigenvalue of the non-viscosity equations. We add (\ref{visc1}) back into the equation (\ref{cases1}). This gives

\begin{equation}
  \begin{cases}
(\partial_t+u_x+u\partial_x )A'=\lambda A', \\ 
(-\partial_t-u\partial_x )A'=\lambda A',
  \end{cases} \label{eigeneq11}
\end{equation}

\noindent
 from which it follows that $A'=A$, and $A$ is the eigenvector of the non-viscosity equations. This means that even in the presence of viscosity, the viscosity does not couple to the eigenvector or it does not change it. We have then
 
 \begin{equation}
  \lambda '
  = \frac{1}{2}u_x- \nu \frac{\partial_{xx}A}{A} .
\label{eq:jac5}
\end{equation}

\noindent
We take the condition (\ref{condit2}) and take a derivative with respect to $x$

\begin{equation}
  uA_{xx} = \frac{3}{4}\frac{u_x^2}{u}A-\frac{1}{2}u_{xx}A, \label{eigeq1212}
\end{equation}

 \noindent
and we get the eigenvalue 
 
  \begin{equation}
  \lambda '= \frac{1}{2}u_x- \nu F[u] = 
  \frac{1}{2}u_x -\nu \Bigl (\frac{3}{4}\frac{u_x^2}{u^2} -\frac{u_{xx}}{2u}. \Bigr ).
\label{eq:jac512}
\end{equation}
 
 \noindent
 The Jacobian is

\begin{equation}
    J[u] = \exp \Bigl (  \int dtdx \ln ( \frac{1}{2}u_x -\nu \Bigl (\frac{3}{4}\frac{u_x^2}{u^2} - \frac{u_{xx}}{2u} \Bigr ) )\Bigr ), \label{det20}
\end{equation}

\noindent
and the effective action then becomes 

\begin{eqnarray}
  S_{eff}[u,\mu] &=&  \frac{1}{2} \int dtdxdy \mu(t,x)\kappa(x-y)\mu(t,y) -i\int
dtdx \mu(u_t+uu_x-\nu u_{xx})  \nonumber \\
&&-\int dtdx \ln ( \frac{1}{2}u_x -\nu \Bigl (\frac{3}{4}\frac{u_x^2}{u^2} -\frac{u_{xx}}{2u}\Bigr ) ). \label{acteff1}
\end{eqnarray}

\noindent
The anomaly-free theory can be obtained with the condition

\begin{equation}
 \frac{1}{2}
 \partial_x \delta u -\nu \delta F[u]=
 \frac{1}{2}
 \partial_x \delta u - \nu \Big ( 
 \frac{3}{2}\frac{u_x}{u^2} \partial_x \delta u
 -\frac{3}{2}\frac{u_x^2}{u^3}\delta u
 -\frac{\partial_{xx}\delta u}{2u}
 +\frac{u_{xx}}{2u^2} \delta u
  \Bigr ) = 0.\label{constr3}
\end{equation}

\noindent
Viscosity becomes part of the constraint.

\section{\label{sec5}Local symmetry}

In this section, we investigate the local symmetries of the effective action (\ref{acteff1}). We  consider the following local time reparameterization, $\beta = \beta(t)$,  $a$ and $b$ are constants,

\begin{eqnarray}
  &&\tilde{t} = \beta(t)^a, \label{local1}\\
  &&\tilde{x} = x\beta'(t)^b ,\\
  &&\tilde{u} = u\beta'(t)^{a-b} ,\\
  &&\tilde{\mu} =\mu\beta'(t)^{2b-a}. \label{local2}
\end{eqnarray} 

\noindent
This translates to the following field variations

\begin{eqnarray}
  &&\delta u = a\beta u_t +b\beta'xu_x+(a-b)\beta' u,
\label{eq:deltau} \\
  &&\delta \mu = a\beta \mu_t +b\beta'x\mu_x+(2b-a)\beta'\mu.\label{eq:deltapsi}
\end{eqnarray}

\noindent
 The transformation is a combination of symmetries (\ref{detd}) and (\ref{detd1}) where 
 $\alpha(t) = a\beta(t)$ and $\epsilon(t,x) = b\beta'(t)x$.  Based on the previous discussion in this article, the assumption is that we should see two local symmetries.  
 We first consider the case where the viscosity is set to zero, $\nu=0$, and then add the viscosity term to the action. \\

\noindent
{\bf Zero viscosity} \\

We apply transformations (\ref{local1})-(\ref{local2})  to action (\ref{acteff110}). The result is 

\begin{equation}
  \delta S_{eff} [\nu = 0]= \frac{2b-3a}{2}\int dtdxdy \beta' \mu(x)\kappa(x-y)\mu(y) 
- i  \int dtdx \beta'' \mu (bxu_x+(a-b)u)  -2\int dtdx \frac{1}{u_x}\partial_x \delta u .
\label{eq:final0}
\end{equation}

\noindent
This can be rewritten as, $b \neq 0$,

\begin{equation}
  \frac{\delta S_{eff}[\nu = 0]}{b} = \frac{3h-1}{2}\int dtdxdy \beta' \mu(x)\kappa(x-y)\mu(y) 
- i  \int dtdx \beta'' \mu (xu_x-hu)-\frac{2}{b}\int dtdx \frac{1}{u_x}\partial_x \delta u ,
\label{eq:final00}
\end{equation}

\noindent
where $h=(b-a)/b$. The second term can vanish either if

\begin{eqnarray}
 &{\rm a)}&\beta'' = 0, \hspace*{2cm}{\rm or}  \label{condi1} \\
 &{\rm b)}&xu_x - hu = C/\beta'', \label{condi2}
\end{eqnarray}

\noindent
where $C$ is a constant. The first condition (\ref{condi1}) links to local symmetry ({\ref{detd}) and the second condition (\ref{condi2}) relates to symmetry ({\ref{detd1}).  We consider the constraint of the anomaly, $\partial_x \delta u = 0$, and make use of (\ref{eq:deltau}). We get

\begin{equation}
  \delta u= a\beta u_t + b \beta' (xu_x-hu) = g(t), \label{condi}
\end{equation}

\noindent
and $g(t)$ is a time-dependent function. If the stochastic Burgers equation has local symmetries, the condition (\ref{condi}) creates a constraint on $u$. We consider two cases based on  (\ref{condi1}}) and (\ref{condi2}).\\

\noindent
a) Case $\beta'' = 0$} \\

Without the anomaly this is a local symmetry for the "bosonic" part of the stochastic Burgers equation (pump term can be parametrized to be zero). 
When $\beta''$ is set to zero, the condition (\ref{condi}) takes the form

\begin{equation}
  \delta u(\beta'' = 0) = a(ct+d) u_t + b c(xu_x-hu) = g(t), \label{condi3}
\end{equation}

\noindent
where $\beta = ct+d$, $c$ and $d$ are constants. When the constraint (\ref{condi3}) is effective, the second term in (\ref{eq:final00}) and the anomaly vanish and what remains in the equation (\ref{eq:final00}) is the pump term. The pump is zero when $h=1/3$. This is Kolmogorov's scaling law for self-similar flows, Ref.~\onlinecite{frisch}. As shown in ~\onlinecite{frisch}, Kolmogorov's four-fifths law also requires $h=1/3$.

A special case is obtained if we further choose $\beta(t) = t$ and $g(t) = 0$, then $\epsilon(t,x) = btx$, and  (\ref{condi3}) becomes

\begin{equation}
  \delta u = a(tu_t + u) + b(xu_x-u) = 0. \label{condi4}
\end{equation}

\noindent
The two local symmetries, (\ref{detd}) and (\ref{detd1}), are coupled in a constraint and enable local symmetry for the full stochastic Burgers equation. \\

\noindent
b) Case $xu_x - hu = C/\beta''$  \\

We further assume a situation when the velocity field is constant, e.g. $u_t = 0$, and (\ref{condi}) reduces to 

\begin{equation}
    xu_x-hu = g(t).
\end{equation}

\noindent
By choosing $g(t) = C/\beta''$, the second term in the effective action (\ref{eq:final00}) becomes 

\begin{equation}
- i C \int dtdx \mu.
\label{eq:final}
\end{equation}

\noindent
This integral is zero based on the conservation of motion of the center of mass, Ref.~\onlinecite{ivaskevich}. Again, the pump term is zero when $h=1/3$, and we have a local symmetry for the full stochastic Burgers equation.\\

\noindent
{\bf Non-zero viscosity} \\

We now apply the transformation (\ref{local1})-(\ref{local2})  to action (\ref{acteff1}). The result is 

\begin{eqnarray}
  \frac{\delta S_{eff}}{b} &=& \frac{3h-1}{2}\int dtdxdy \beta' \mu(x)\kappa(x-y)\mu(y) 
- i  \int dtdx \beta'' \mu (xu_x-hu) \nonumber \\
&&
+ i (h+1)\nu \int dtdx \beta'\mu u_{xx}  -\int dtdx \Bigl (\frac{1}{2} u_x - \nu  F[u]  \Bigr ) ^{-1}
  \Bigl ( \frac{1}{2}\partial_x \delta u -\nu \delta F[u]  \Bigr ) .
\label{eq:final000}
\end{eqnarray}

\noindent
In a same way, by setting

\begin{equation}
 \beta''=0, \hspace*{1cm}{\rm or}\hspace*{1cm}xu_x-hu = C/\beta'',
\end{equation}

\noindent
we can observe a local symmetry.
We choose the constraints $\partial_x \delta u = 0$, $u_t = 0$ and $xu_x-hu = C/\beta''$. This is analogous to "fixing a gauge" or "selecting a vacuum". The effective action becomes

\begin{eqnarray}
  \frac{\delta S_{eff}}{b} &=& \frac{3h-1}{2}\int dtdxdy \beta' \mu(x)\kappa(x-y)\mu(y) 
+ i (h+1)\nu \int dtdx \beta'\mu u_{xx} \nonumber \\
&& +\nu C\int dtdx \frac{\beta'}{\beta''}  \Bigl (\frac{1}{2} u_x - \nu F[u]  \Bigr )  ^{-1}
  \Bigl ( \frac{-3u_x^2+uu_{xx}}{2u^3}
\Bigr ).
\label{eq:finalf}
\end{eqnarray}

\noindent
The equation is balanced; the pump term produces energy and the viscosity terms dissipate it. The pump term vanishes at $h = 1/3$ and the other term vanishes at $h = -1$. These cases are explained in more detail in Ref.~\onlinecite{frisch}. The last viscosity term comes from the determinant. The "gauge" could also be selected differently, for example so that the anomaly disappears completely, Eq.(\ref{constr3}). Here we chose the constraint to give the same non-invariance as in Eq.(\ref{det1a11b}).

\section{\label{sec6}Conclusions}

In this article, we have studied the local symmetries of the Burgers equation. Examining the determinant of the equation, we found an anomaly that creates additional structure to the theory. By requiring the anomaly to disappear, the velocity field gains constraints. Using these constraints, we can show two local symmetries applicable to Burgers turbulence. This is also known as Kolmogorov's second law of self-similarity. There is a lot of physics in the determinant. More interesting observations can be made by applying the techniques and methods of field theories.

\end{document}